\documentclass{article}

\usepackage{PRIMEarxiv}

\usepackage[utf8]{inputenc} 
\usepackage[T1]{fontenc}    
\usepackage{hyperref}       
\usepackage{url}            
\usepackage{booktabs}       
\usepackage{amsfonts}       
\usepackage{nicefrac}       
\usepackage{microtype}      
\usepackage{lipsum}
\usepackage{fancyhdr}       
\usepackage{graphicx}       
\graphicspath{{media/}}     
\usepackage{mathtools}
\usepackage{amsmath}
\usepackage{amssymb}
\usepackage{commath}
\usepackage{graphicx}
\usepackage{makecell}

\title{A systematic literature review on solution approaches for the index tracking problem in the last decade
}

\author{
  Julio Cezar Soares Silva, Adiel Teixeira de Almeida Filho \\
  Centro de Informática \\
  Universidade Federal de Pernambuco \\
  Recife\\
  \texttt{\{jcss4,adielfilho\}@cin.ufpe.br} \\   
}

\begin{document}
\maketitle

\begin{abstract}
The passive management approach offers conservative investors a way to reduce risk concerning the market. This investment strategy aims at replicating a specific index, such as the NASDAQ Composite or the FTSE100 index. The problem is that buying all the index's assets incurs high rebalancing costs, and this harms future returns. The index tracking problem concerns building a portfolio that follows a specific benchmark with fewer transaction costs. Since a subset of assets is required to solve the index problem this class of problems is NP-hard, and in the past years, researchers have been studying solution approaches to obtain tracking portfolios more practically. This work brings an analysis, spanning the last decade, of the advances in mathematical approaches for index tracking. The systematic literature review covered important issues, such as the most relevant research areas, solution methods, and model structures. Special attention was given to the exploration and analysis of metaheuristics applied to the index tracking problem.
\end{abstract}


\section{Introduction}
Investment analysis is a strategic process for an organization and individual investors as it involves capital allocations that need to be made efficiently. In this scenario, financial agents constantly deal with the tradeoff between walking through a riskier path (under the perspective of higher returns) and taking less risk and being satisfied with lower (but less uncertain) returns.
Since the seminal studies by \cite{Markowitz1952} and \cite{Roy1952}, which represented the starting points of Modern Portfolio Theory, several papers have been published and advances have been observed in the portfolio selection problem, as shown in \cite{Kolm2014}. Kolm et al. \cite{Kolm2014} affirm that financial experts are apprehensive about the application of the classical MVO in real data, especially because of the sensitivity of the optimal weight allocation relative to the perturbation of the model inputs. Also, the generalized difficulty intrinsic to parameter estimation, as discussed in \cite{Fama1970}, being impossible to predict future returns one or more days before the portfolio's rebalancing day, brings more difficulties concerning the consistency and robustness of results produced using this model.

In this way, the result of the optimization using this historical data approach may lead to counter-intuitive portfolios, therefore more robust models are needed \cite{fabozzi2007,Han2022, Ferreira2021}. Some of the challenges of this field are associated with approaches to model the information uncertainty and ambiguity \cite{Lotfi2018,yoshida2020,cheng2021,guo2021}, stock price prediction \cite{Dutta2018, ricardo2018, ricardo2019, Thakkar2021}, incorporation of constraints that reflect investor's preferences and practical issues related to the market \cite{ferreira2018, Silva2021, silva_imodrsa_2021, delimasilva2020, deLimaSilva2018,Pasricha2020,Kouaissah2020}, cryptocurrencies \cite{Li2021}, and new paradigms, such as portfolio selection based on assets networks \cite{puerto2020,clemente2019}, represent some of the challenges that researchers face.


A passively managed fund is known as \textit{index fund/tracker fund}. A manager that adopts this strategy could buy all the stocks of a given stock index and reproduce it perfectly (full replication), but this strategy has some disadvantages \cite{beasley2003,Canakgoz2009, santanna2017_heur}:

\begin{itemize}
    \item The composition of the index is revised periodically. Therefore, the holdings of all stocks will change periodically to reflect the new composition's weights of the index.
    \item Transaction costs associated with the index's stocks cannot be limited since it is necessary to trade all stocks to reduce the tracking error periodically. 
\end{itemize}

The index tracking problem is concerned with index replication, but limiting transaction costs by using fewer stocks. Index tracking models are classified as combinatorial optimization problems when they integrate cardinality constraints in other to control the number of assets, such that the costs are reduced by using a small number of stocks (compared to the number of constituents of the index to be replicated) in the portfolio \cite{beasley2003,gaivoronski2005,santanna2017_heur}. 

This problem is difficult for a computer to solve as the number of stocks grows. The majority of combinatorial optimization problems are NP-complete and NP-hard. Some solution approaches are exact algorithms, such as branch-and-cut and heuristics, which consists of algorithms that provide a feasible solution without quality and computational time guarantees \cite{papadimitriou1998, marti2018, resende2016}.

Since the study of Beasley et al. \cite{beasley2003}, which is a popular study considering an approximate solution approach for index tracking, many computational studies have been developed for this kind of problem and portfolio selection. Recent studies concerning index tracking were developed mainly by the popularization of metaheuristics/heuristics as tools to search for approximate optimal solutions \cite{andriosopoulos2019}, and also the importance of considering rebalancing process strategies \cite{STRUB2018, santanna2017_heur}.

We present a literature review comprising solution approaches for the index tracking problem over the last decade. The systematic study raises relevant questions such as most used model structures, most used solution approaches, most relevant solution methods and most used data sources. This brings an overview of the development of index tracking research over the last decade.



\section{Review methodology}

The procedures and rules that conducted the systematic literature review were based on \cite{kalayci2019} and \cite{almeida-filho2021}. \cite{kalayci2019} presented a comprehensive literature review about solution approaches for the portfolio selection problem guided by research questions.
\cite{almeida-filho2021} also used research questions to develop a systematic literature review that encompasses financial modelling with multicriteria decision making.

\subsection{Research questions set up}
To guide the result analysis and scope of this work, some research questions were defined.The research questions are presented in Table \ref{tab:rqs_idx}

\begin{table*}[h!]
\centering
\scalebox{0.9}{
\begin{tabular}{cc}
\hline
RQ  & Description                                                                                                                                   \\ \hline

\#1 & \makecell{Are index tracking solution methods more relevant to journals focusing \\on operations research and computer science? }                                                            \\
\#2 & \makecell{Is there a concentration of heuristic methods applied in a specific\\quantitative modelling framework?} 

\\
\#3 & \makecell{Has there been a growth in the number of non-heuristic methods applied to \\ the index tracking problem?}                                                            \\
\#4 & \makecell{Has there been a growth in the number of heuristics/metaheuristics applied \\ to the index tracking problem?}                                                           \\
\#5 & \makecell{Are heuristic approaches more used than non-heuristic approaches for \\index tracking problems? }                                                            \\
\#6 & \makecell{Do heuristic approaches have more cite impact than non-heuristic \\ approaches for index tracking problems? }   
\\
\#7 & \makecell{Is there a prevalence of using a specific heuristic/metaheuristic \\in index tracking problems?}
\\
\#8 & \makecell{Is there an integration between heuristic and general-purpose solvers?}
\\
\#9 & \makecell{Is there a prevalence of using specific evaluation metrics for heuristic\\ approaches?}
\\
\#10 & \makecell{Is there a prevalence of using a specific solution method to \\ compare with heuristic approaches?}
\\
\#11 & \makecell{Is there a prevalence of solving for a specific tracking error objective \\function when using heuristic approaches?}
    \\
\#12 & \makecell{Is there a prevalence of solving for specific practical constraints when\\ using heuristic approaches?}
    \\
\#13 & \begin{tabular}[c]{@{}c@{}}Which databases were most adopted in heuristic approaches? 
    \end{tabular} \\ \hline
\end{tabular}
}
\caption{Research questions for index tracking systematic literature review}
\label{tab:rqs_idx}
\end{table*}

The research questions related to important sources, production, citation impact (RQs 1-6) were developed based on the work of \cite{almeida-filho2021}. The research questions related to the heuristic approaches (RQs 7-13) were developed taking into consideration the work of \cite{kalayci2019}

\subsection{Material collection and selection}

The search was conducted by covering articles and proceedings papers published from 2010 to 2021. The bibliography collection process took place at the Web of Science database using the keyword structure presented in Table \ref{tab:keywordsidx}. This keyword combination structure was based on \cite{kalayci2019} and keywords related to index tracking were included from the author's a priori knowledge of the field. Levels 1 and 2 guide the search for financial portfolio optimization problems. Level 3 restricts the financial portfolio problem class to index tracking.

\begin{table*}[h!]
\centering
\begin{tabular}{@{}cc@{}}
\toprule
Level & Search Terms                                                                                         \\ \midrule
1     & \begin{tabular}[c]{@{}c@{}}Portfolio OR Investment OR Asset\\ AND\end{tabular} \\
2     & \begin{tabular}[c]{@{}c@{}}Optimization OR Management OR Selection\\ AND\end{tabular}        \\
3     & "Index Tracking" OR "Tracking Error" OR Tracking-Error\\  
\bottomrule
\end{tabular}
\caption{Proposed keyword combination structure for the index tracking systematic review}
\label{tab:keywordsidx}
\end{table*}

\begin{figure*}[h!]
    \centering
    \includegraphics[width=\linewidth]{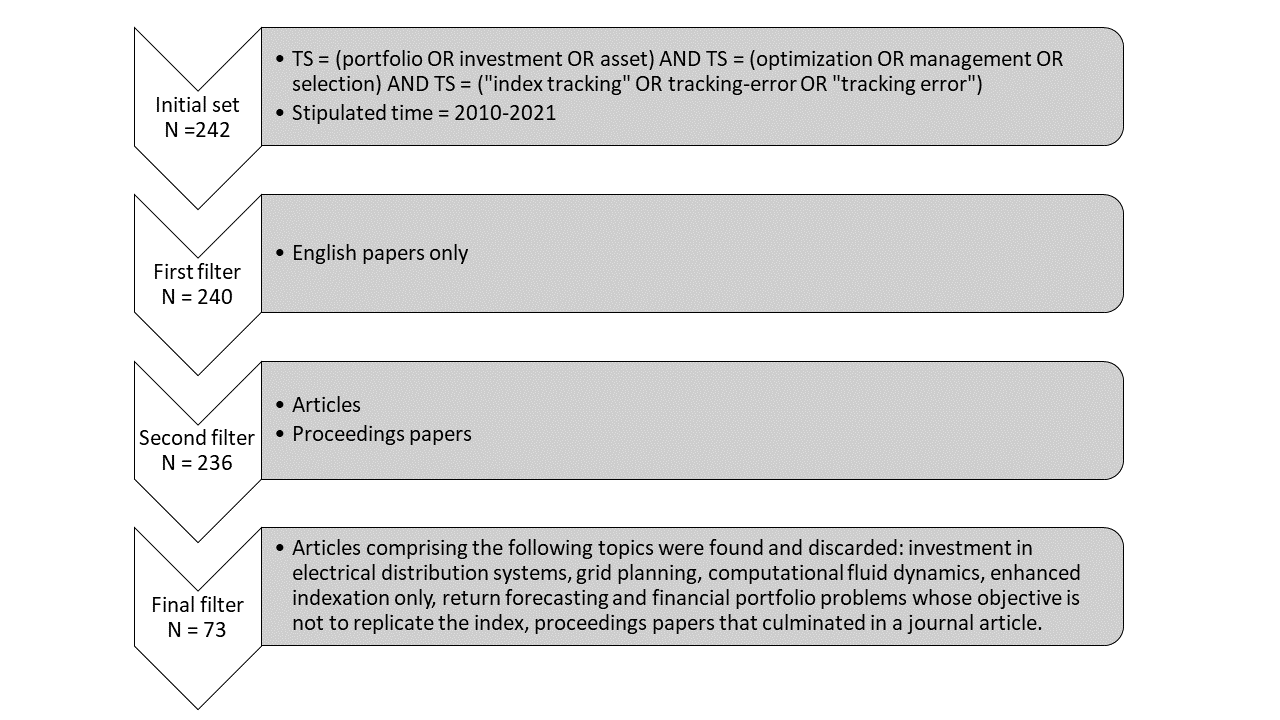}
 \caption{Filtering process of the index tracking literature review}
 \label{fig:filter_litrev}
\end{figure*}

The filtering process is shown in Figure \ref{fig:filter_litrev}. 
The first filter was used to select only papers that were written in the English language. The second filter was applied to limit the type of documents to articles and proceedings papers. The final filter was performed by fully reading the papers and selecting those that fitted the scope of this work. Many of the outs of the scope works appeared because of the first and second levels keywords. Works that were eliminated did not approach financial portfolio optimization problems or were portfolio optimization problems whose objective was not to replicate the index, or proceedings papers that culminated in journal articles.

A total of 73 articles were selected after applying all the filters. Of the 73 papers, 64 are journal articles and 9 are conference papers. Some research questions needed the support of bibliometric tools to be answered. Bibliometrix \cite{bibliometrix} and Web Of Science were adopted to perform these analyses.

\section{Review results}

The answer to RQ\#1 was constructed by considering scientific journals that are responsible for publications involving index tracking applications. Table \ref{tab:sources1} presents the percentage of sources according to their associated number of publications.

\begin{table*}[h!]
\centering
\begin{tabular}{@{}cccc@{}}
\toprule
Publications & Number of sources & (\%)Journals  & (\%) Accumulated                                                             \\ \midrule
1     & 40  &  78.43  &   78.43 \\
2     & 6   &   11.76   &  90.19 \\
3     & 2   &   3.92    &   94.11 \\
$>$3     & 3   &   5.88    &  100.00\\
\midrule
\textbf{Total} & 51 & 100 &  \\
\bottomrule
\end{tabular}
\caption{Percentage of sources according to their total number of index tracking publications}
\label{tab:sources1}
\end{table*}

The majority of paper sources (40), which represents 78.43\% of the total sources, are responsible for only one publication. This result shows that, despite being new literature, where the first research papers developed by \cite{consiglio2001} and \cite{konno2001}, index tracking papers are scattered in a wide range area spectrum, such as economy, statistics, computer science and operations research. A more specific analysis was developed by considering 11 journals that had published more than one article (21.57\% of the total sources) in the considered period. The total number of articles published on the selected sources is equal to 33, which comprises about 45\% of the total published papers in the last decade. This shows that these 11 journals concentrate a good part of the total publications concerning index tracking in the last decade. Table \ref{tab:sources2} shows the name of the source, the number of publications, the relative percentage of publications and the associated Web of Science research category.  

\begin{table*}[h!]
\centering
\scalebox{0.8}{
\begin{tabular}{@{}cccc@{}}
\toprule
Name & Publications & \% of 33  & Categories                                                            \\ \midrule
\makecell{Quantitative Finance}     & 7  &  21.21  &   \makecell{Business, Economics\\, Mathematics}\\ \\
\makecell{Annals of Operations\\Research}     & 4  &  12.12  &   \makecell{Operations Research}\\ \\
\makecell{European Journal of\\Operational Research}     & 4  &  12.12  &   \makecell{Operations Research}\\ \\
\makecell{Computers \& Operations\\Research}     & 3  &  9.09  &   \makecell{Computer Science, \\Engineering, and \\Operations Research}\\ \\
\makecell{Journal of the Operational\\Research Society}     & 3  &  9.09  &   \makecell{Management, and \\Operations Research}\\ \\
\makecell{Applied Soft Computing}     & 2  &  6.06  &   \makecell{Computer Science}\\ \\
\makecell{Computational Statistics \&\\Data Analysis}     & 2  &  6.06  &   \makecell{Computer Science,\\ and Statistics \& Probability}\\ \\
\makecell{Expert Systems with \\Applications}     & 2  &  6.06  &   \makecell{Computer Science, \\Engineering, and Operations \\Research}\\ \\
\makecell{Journal of Economic\\Dynamics \& Control}     & 2  &  6.06  &   \makecell{Economics}\\ \\
\makecell{Journal of Risk and \\Financial Management}     & 2  &  6.06  &   \makecell{Business}\\
\makecell{Optimization Letters}     & 2  &  6.06  &   \makecell{Operations Research, and\\Mathematics}\\
\bottomrule
\end{tabular}
}
\caption{Selected sources}
\label{tab:sources2}
\end{table*}

The following Web of Science research categories were reached: Operations Research (including Operations Research \& Management), Management, Business (including Business \& Finance), Statistics \& Probability, Economics, Mathematics (including Mathematical Methods in Social Sciences and Applied Mathematics), Computer Science (Interdisciplinary and Artificial Intelligence), and Engineering (including Industrial, and Electrical \& Electronic). Table \ref{tab:sources3} contains the number of journals and publications associated with each of the categories that were found. The answer to RQ \#1 is positive since the most relevant journals cover operations research and computer science research areas. Computer Science contains the same number of related publications in Economics, Mathematics and Business categories, but it scored better because its related publications are distributed among 4 journals with better impact factors.

\begin{table*}[h!]
\centering
\begin{tabular}{@{}ccc@{}}
\toprule
Category & \makecell{Number of related\\journals} &
\makecell{Number of related\\publications}                                                            \\ \midrule
Operations Research &   6   &   18  \\
Computer Science    &   4  & 9   \\
Economics   & 2 &   9  \\
Mathematics     &   2   &   9\\
Business   &   2   &   9  \\
Engineering  &   2   &   5 \\
Management  & 1    &    3  \\
Statistics \& Probability  &   1   &   3 \\
\bottomrule
\end{tabular}
\caption{The relevant categories}
\label{tab:sources3}
\end{table*}

\subsection{Categorization - Quantitative modelling frameworks and solution approaches}

To answer RQs \#2, \#3, \#4, \#5 and \#6 it was necessary to categorize the modelling framework first and then categorize the solution approach. Three categories of quantitative modelling frameworks were defined: mathematical programming frameworks, statistical techniques frameworks, and other frameworks. This taxonomy was constructed based on the models used in the selected works. Tables \ref{tab:model_frmwks_mp}, \ref{tab:model_frmwks_stat}, and \ref{tab:model_frmwks_other}, shows the publications associated with the Mathematical modelling, statistical techniques, and other frameworks categories, respectively.

\begin{table}[h!]
\centering
\scalebox{0.95}{
\begin{tabular}{@{}cc@{}}
\toprule
Subcategory & Publications                                                                     \\ \midrule

Linear Programming
& \multicolumn{1}{l}{
\begin{tabular}[c]{@{}l@{}}
\scalebox{0.8}{\cite{goel2018}; }
\scalebox{0.8}{\cite{kusiak2013};}\\
\scalebox{0.8}{\cite{theobald}}\\
\end{tabular}
} \\ \\
\makecell{Mixed-integer\\linear programming }
& \multicolumn{1}{l}{
\begin{tabular}[c]{@{}l@{}}
\scalebox{0.8}{\cite{chen2012}; }
\scalebox{0.8}{\cite{goel2018};}\\
\scalebox{0.8}{\cite{guastaroba_2012}; }
\scalebox{0.8}{\cite{huang2018};}\\
\scalebox{0.8}{\cite{mezali2013}; }
\scalebox{0.8}{\cite{mezali2014};}\\
\scalebox{0.8}{\cite{wang2012_main}; }
\scalebox{0.8}{\cite{wu2017_clust};}\\
\scalebox{0.8}{\cite{strub2015_proceedings}}\\
\end{tabular}
} \\ \\
Non-linear Programming
& \multicolumn{1}{l}{
\begin{tabular}[c]{@{}l@{}}
\scalebox{0.8}{\cite{penev2021}; }
\scalebox{0.8}{\cite{glabadanidis2020}}
\end{tabular}
} \\ \\
\makecell{Mixed-integer\\Non-linear programming }
& \multicolumn{1}{l}{
\begin{tabular}[c]{@{}l@{}}
\scalebox{0.8}{\cite{andriosopoulos2013};}\\
\scalebox{0.8}{\cite{andriosopoulos2014};}\\
\scalebox{0.8}{\cite{garcia2018};}
\scalebox{0.8}{\cite{fastrich2012};}\\
\scalebox{0.8}{\cite{grishina2017}; } 
\scalebox{0.8}{\cite{huang2018};}\\
\scalebox{0.8}{\cite{mutunge_2018}; }
\scalebox{0.8}{\cite{ni2013};} \\
\scalebox{0.8}{\cite{santanna2017_heur};}
\scalebox{0.8}{\cite{santanna2017_exact};}\\
\scalebox{0.8}{\cite{santanna2019}; }
\scalebox{0.8}{\cite{scozzari_2013};}\\
\scalebox{0.8}{\cite{strub2019}; }
\scalebox{0.8}{\cite{valle2015};}\\
\scalebox{0.8}{\cite{wang_2018}; }
\scalebox{0.8}{\cite{xu2016};}\\
\scalebox{0.8}{\cite{strub2016_proceedings}; \cite{vieira2021};}\\
\scalebox{0.8}{\cite{santanna2020_2}; \cite{fernandez-lorenzo2021};}\\
\scalebox{0.8}{\cite{affolter2016}}
\end{tabular}
} \\ \\
Conic Programming
& \multicolumn{1}{l}{
\begin{tabular}[c]{@{}l@{}}
\scalebox{0.8}{\cite{ling2014}; }
\end{tabular}
} \\ \\
\makecell{Mixed-integer\\Conic Programming}
& \multicolumn{1}{l}{
\begin{tabular}[c]{@{}l@{}}
\scalebox{0.8}{\cite{wu2019}}
\end{tabular}
} \\ \\
Stochastic programming 
& \multicolumn{1}{l}{
\begin{tabular}[c]{@{}l@{}}
\scalebox{0.8}{\cite{barro2014};}\\
\scalebox{0.8}{\cite{barro2019};}
\scalebox{0.8}{\cite{stoyan2010}}\\
\end{tabular}
} \\ \\
Dynamic programming 
& \multicolumn{1}{l}{
\begin{tabular}[c]{@{}l@{}}
\scalebox{0.8}{\cite{cheng2013}; }
\scalebox{0.8}{\cite{stoyan2010}}\\
\end{tabular}
} \\ \\
\makecell{Multi-Objective\\Optimization}
& \multicolumn{1}{l}{
\begin{tabular}[c]{@{}l@{}}
\scalebox{0.8}{\cite{bilbao_terol2012};}\\
\scalebox{0.8}{\cite{chiam2013}; }
\scalebox{0.8}{\cite{garcia2011};}\\
\scalebox{0.8}{\cite{garcia2013}; }
\scalebox{0.8}{\cite{li2014};}\\ 
\scalebox{0.8}{\cite{ni2013}; }
\scalebox{0.8}{\cite{wu2014_fuzzy}; }\\
\scalebox{0.8}{\cite{zhang2018_proceedings}}
\end{tabular}
} \\ 
\bottomrule
\end{tabular}
}
\caption{Publications associated with the mathematical programming category}
\label{tab:model_frmwks_mp}
\end{table}

\begin{table}[h!]
\centering
\scalebox{1.2}{
\begin{tabular}{@{}cc@{}}
\toprule
Subcategory & Publications                                                                     \\ \midrule
Cointegration
& \multicolumn{1}{l}{
\begin{tabular}[c]{@{}l@{}}
\scalebox{0.8}{\cite{acosta-gonzalez2015}; }\\
\scalebox{0.8}{\cite{papantonis2016}; }
\scalebox{0.8}{\cite{santanna2017_exact};}\\
\scalebox{0.8}{\cite{santanna2019}; }
\scalebox{0.8}{\cite{santanna2020}}
\end{tabular}
} \\ \\
\makecell{Regression}
& \multicolumn{1}{l}{
\begin{tabular}[c]{@{}l@{}}
\scalebox{0.8}{\cite{siew2016}; }
\scalebox{0.8}{\cite{chen2011_proceedings}}
\end{tabular}
} \\ \\
\makecell{Regression with\\regularization}
& \multicolumn{1}{l}{
\begin{tabular}[c]{@{}l@{}}
\scalebox{0.8}{\cite{benidi2018}; }
\scalebox{0.8}{\cite{fastrich2012}}\\
\scalebox{0.8}{\cite{giuzio2016}; }
\scalebox{0.8}{\cite{shu2020};}\\
\scalebox{0.8}{\cite{giuzio2017}; }
\scalebox{0.8}{\cite{tas2018}; }
\scalebox{0.8}{\cite{xu2016}} \\
\end{tabular}
} \\ \\
\makecell{Regression with\\ regularization and\\variable selection} 
& \multicolumn{1}{l}{
\begin{tabular}[c]{@{}l@{}}
\scalebox{0.8}{\cite{wu2014_elastic}; }
\scalebox{0.8}{\cite{wu2014_lasso};}\\
\scalebox{0.8}{\cite{yang2016}; }
\scalebox{0.8}{\cite{zhao2016}; }
\scalebox{0.8}{\cite{li2020};} \\
\scalebox{0.8}{\cite{santanna2020}}
\end{tabular}
} \\ 
\bottomrule
\end{tabular}
}
\caption{Publications associated with the statistical techniques category}
\label{tab:model_frmwks_stat}
\end{table}

\begin{table}[h!]
\centering
\scalebox{1.2}{
\begin{tabular}{@{}cc@{}}
\toprule
Subcategory & Publications                                                                     \\ \midrule
Machine Learning
& \multicolumn{1}{l}{
\begin{tabular}[c]{@{}l@{}}
\scalebox{0.8}{\cite{ouyang2019}; }
\scalebox{0.8}{\cite{nakayama2018};}\\
\scalebox{0.8}{\cite{kim2020}; }
\scalebox{0.8}{\cite{ni2013_proceedings}; }
\scalebox{0.8}{\cite{tang2014_proceedings}; } \\
\scalebox{0.8}{\cite{strub2015_proceedings}; \cite{li2011};}\\
\scalebox{0.8}{\cite{zhang2020}; \cite{hong2021};} \\
\scalebox{0.8}{\cite{kwak2021} } \\
\end{tabular}
} \\ \\
Sampling
& \multicolumn{1}{l}{
\begin{tabular}[c]{@{}l@{}}
\scalebox{0.8}{\cite{djoko2015}; } 
\scalebox{0.8}{\cite{boda2016_proceedings}}
\end{tabular}
} \\ 
\bottomrule
\end{tabular}
}
\caption{Publications associated with the other framework category}
\label{tab:model_frmwks_other}
\end{table}

\begin{figure*}[h!]
\centering
    \scalebox{0.75}{
    \includegraphics[width=\linewidth]{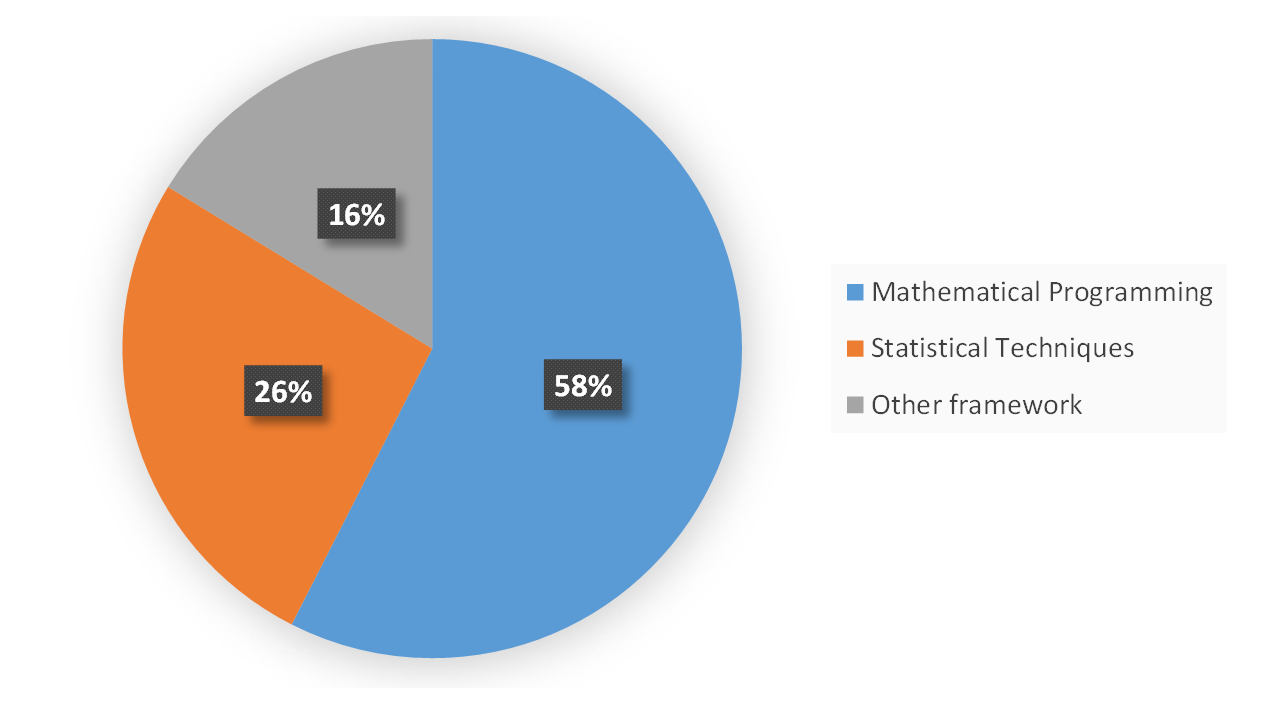}
    }
 \caption{Percentage of works that used each modelling framework}
 \label{fig:modfrmwks}
\end{figure*}

Figure \ref{fig:modfrmwks} depicts the relative percentage of published articles per quantitative modelling framework. It can be observed that the majority of works on the index tracking problem are exploring more mathematical programming frameworks than any other type of modelling framework for this problem. In this work, we consider that heuristic approaches are those that combined stochastic algorithms, such as genetic algorithms, with at least one of the explored quantitative modelling frameworks to solve the index tracking problem. The total number of articles allocated to the heuristic group and non-heuristic is 25 and 48, respectively. Thus, the majority of works on index tracking adopted non-heuristic solution methods. As discussed in \cite{graham2020} an exact algorithm can be efficiently applied to solve small-cardinality portfolios (when it contains less than five assets) or as the cardinality of the portfolio approximates the size of the considered universe of assets. Depending on the size of the portfolio to be optimized, there are specific conditions where heuristics are still more efficient \cite{graham2020}. However, in some specific situations, exact algorithms may be deployed efficiently. 

\subsection{Comparison of production and impact of the solution approaches}

\begin{figure*}[h!]
\centering
    \scalebox{0.75}{
    \includegraphics[width=\linewidth]{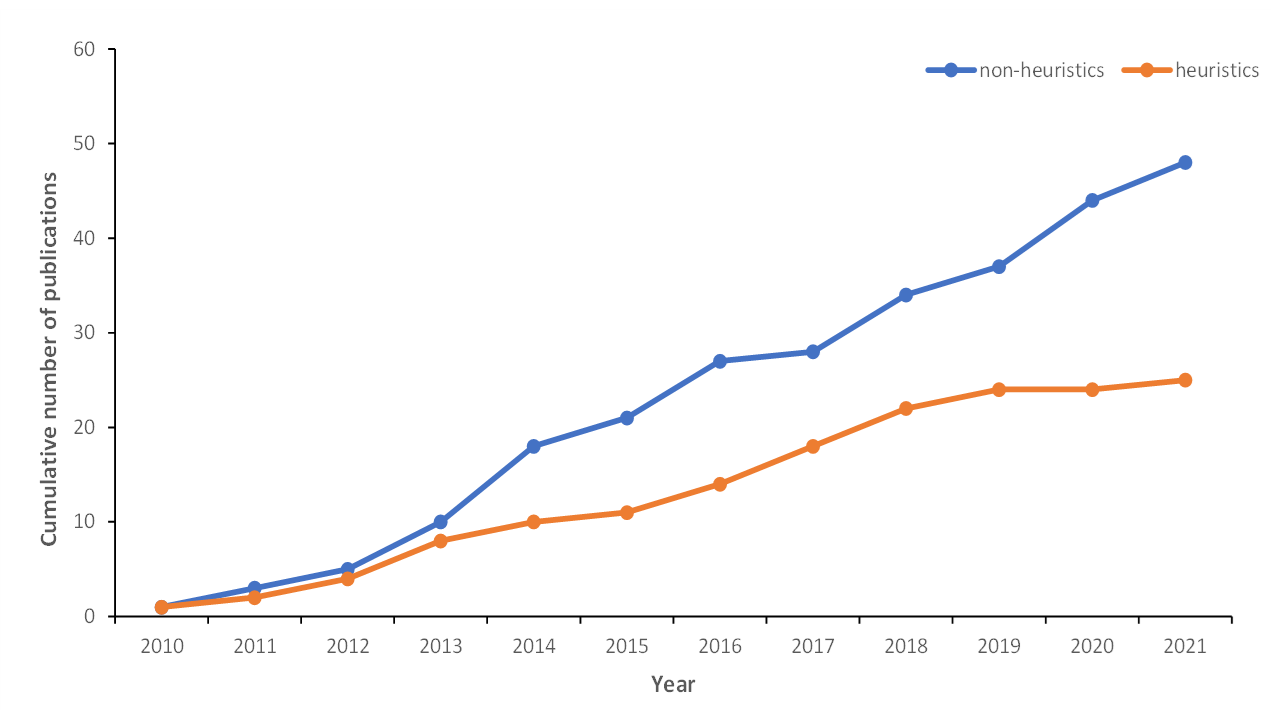}
    }
 \caption{Comparison of cumulative annual total publications for papers using heuristic and non-heuristic approaches}
 \label{fig:annTP}
\end{figure*}

After associating each article with its solution approach, an analysis of production growth and citation impact was developed. Figure \ref{fig:annTP} shows the cumulative number of publications per year for each solution approach. 
Since there has been a growth in the number of non-heuristic and heuristic methods applied to the index tracking problem, the answer for both RQ \#3 and \#4 is positive. The answer to RQ \#5 is negative since the production rate of papers adopting non-heuristic approaches is bigger than that of papers adopting heuristic approaches.

\begin{figure*}[h!]
\centering
\scalebox{0.75}{
    \includegraphics[width=\linewidth]{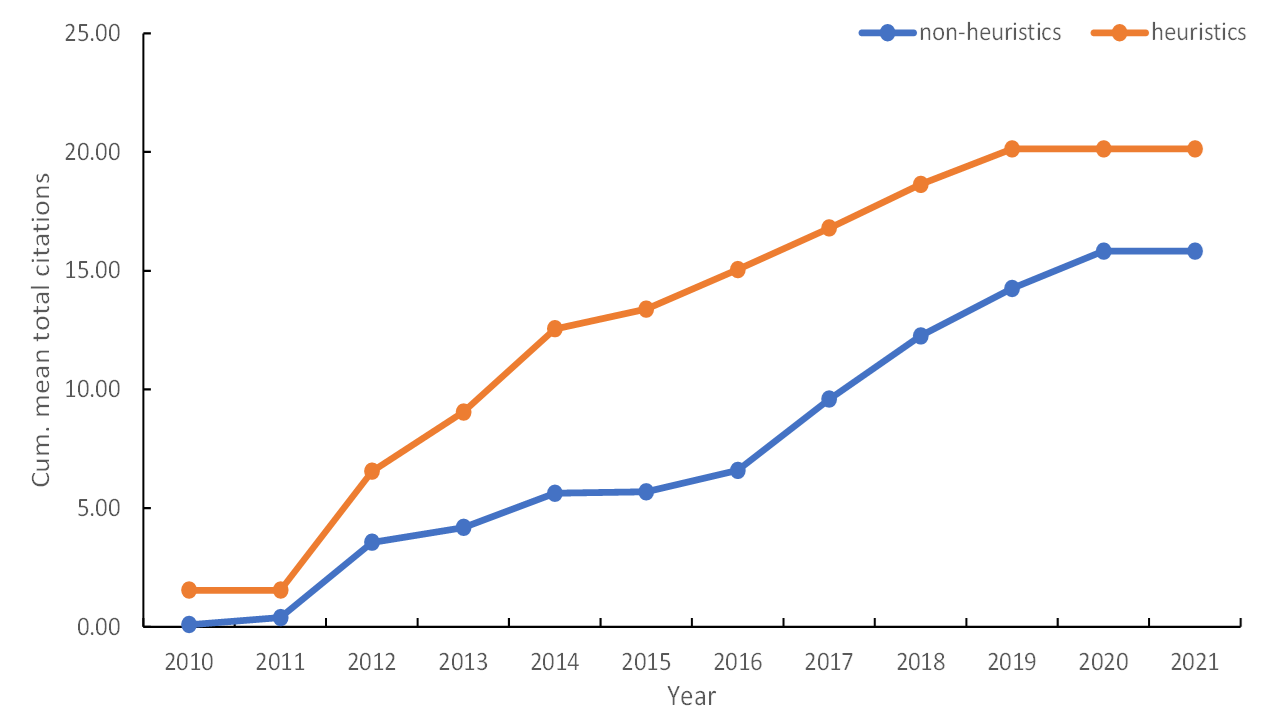}}
 \caption{Comparison of cumulative annual total mean citations for heuristic and non-heuristic approaches}
 \label{fig:annTMC}
\end{figure*}

\begin{figure*}[h!]
\centering
    \scalebox{0.75}{
    \includegraphics[width=\linewidth]{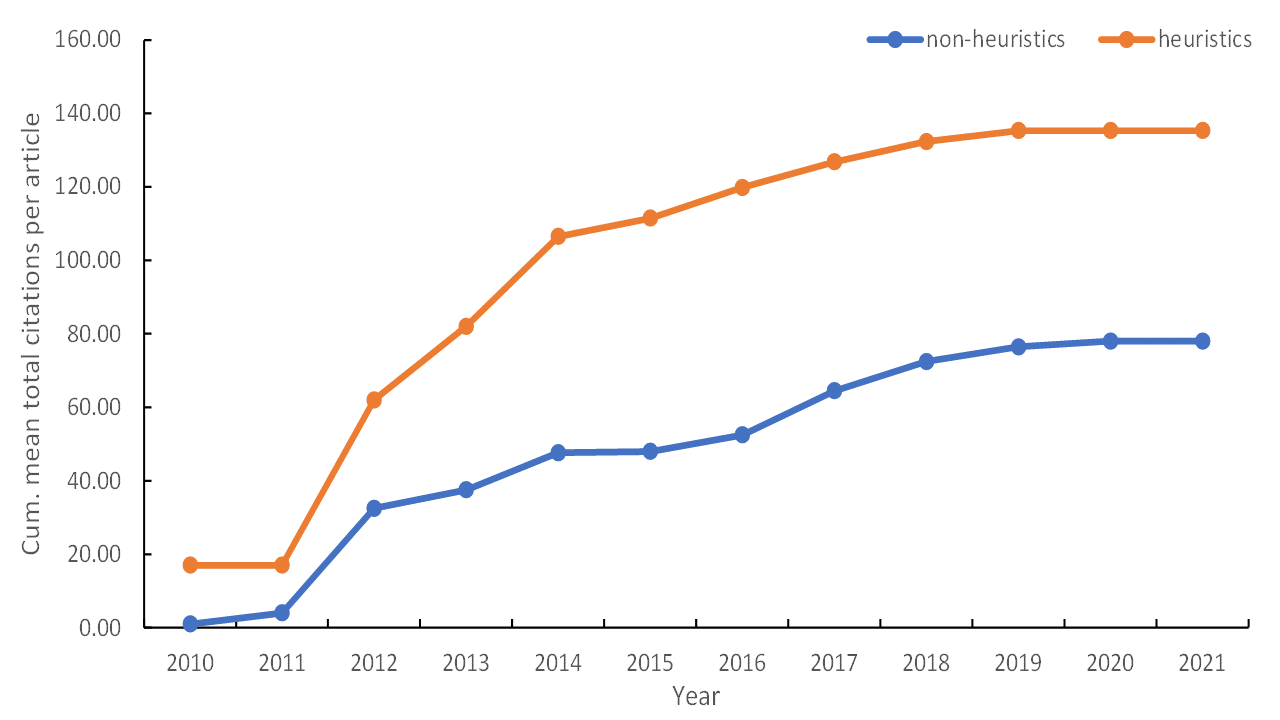}}
 \caption{Comparison of cumulative annual total mean citations per article for heuristic and non-heuristic approaches}
 \label{fig:annTCperArt}
\end{figure*}

Two metrics were evaluated to answer RQ \#6: cumulative annual mean total citations and cumulative annual mean total citations per article. Figure \ref{fig:annTMC} show the cumulative annual total mean citations for each solution approach and Figure \ref{fig:annTCperArt} shows the cumulative annual mean total citations per article.

Papers that adopted heuristic approaches obtained a total of 322 citations over the last decade. The non-heuristic group obtained a total of 281 citations. Yearly mean total citations of heuristic solution approach overcome that of non-heuristic solution approaches in most years. This result is also observed for the annual total mean citations per article metric. The answer to RQ \#6 is positive because the heuristic group surpassed the non-heuristic group in the two evaluation metrics. An interesting observation concerning the production and citation impact result is that although the production of heuristic papers is lower, their impact is higher.

\subsection{Categorization - adopted heuristics/metaheuristics}

Exploratory work was made to find all heuristics used in each of the 25 articles to answer RQs \#2 and \#7. Also, an investigation concerning the hybridization of the heuristics with general-purpose solvers was performed to answer RQ \#8. The heuristics found were: Genetic Algorithm (GA), Differential Evolution (DE), Tabu Search (TS), Greedy Construction (GC), Best Exchange by one (BEBO), Combinatorial Search (CS), Local Branching (LB), Lagrangian-based (LGR), Kernel Search (KS), Variable Neighborhood Search (VNS), Decomposition Algorithm (DA), Multi-objective Genetic Algorithm (MOGA),  Multi-objective Particle Swarm (MOPS), Heuristic Prunning (HP), Simulated Annealing (SA), and Invasive Weed Optimization (IWO). In Table \ref{tab:cat_heur} the performed allocation of papers to a heuristic is shown.

\begin{table}[h!]
\centering
\scalebox{1.3}{
\begin{tabular}{@{}ccc@{}}
\hline
 & Not Hybrid & Hybrid 
\\ \hline
GA   
&  
\makecell{
\scalebox{0.5}{\cite{acosta-gonzalez2015}; }\\
\scalebox{0.6}{\cite{andriosopoulos2013}; }\\
\scalebox{0.6}{\cite{andriosopoulos2014}; }\\
\scalebox{0.6}{\cite{fastrich2012}; }\\
\scalebox{0.6}{\cite{garcia2018}; }\\
\scalebox{0.6}{\cite{giuzio2017}; }
\scalebox{0.6}{\cite{grishina2017}; }\\
\scalebox{0.6}{\cite{ni2013}; }\\
}
& 
\makecell{
\scalebox{0.6}{\cite{santanna2017_heur}; }
\scalebox{0.6}{\cite{wang2012_main}; }\\
\scalebox{0.6}{\cite{wang_2018}; }
\scalebox{0.6}{\cite{xu2016}; }\\
\scalebox{0.6}{\cite{strub2019};}\\
\scalebox{0.6}{\cite{strub2016_proceedings}; }
\scalebox{0.6}{\cite{chen2011_proceedings}; }
}
\\\hline
DE
&  
\makecell{
\scalebox{0.6}{\cite{andriosopoulos2013}; }\\
\scalebox{0.6}{\cite{andriosopoulos2014}; }\\
\scalebox{0.6}{\cite{grishina2017}; }
}
& 
\makecell{
\scalebox{0.6}{\cite{scozzari_2013} }
}
\\\hline
TS 
&  
\makecell{
\scalebox{0.6}{\cite{garcia2018}; }
}
& 
\makecell{
\scalebox{0.6}{ }
}
\\\hline
GC
&  
\makecell{
\scalebox{0.6}{\cite{mutunge_2018}; }\\
\scalebox{0.6}{\cite{strub2016_proceedings}; }
}
& 
\makecell{
\scalebox{0.6}{}
}
\\\hline
BEBO
&  
\makecell{
\scalebox{0.6}{\cite{mutunge_2018}; }
}
& 
\makecell{
\scalebox{0.6}{ }
}
\\\hline
CS
&  
\makecell{
\scalebox{0.6}{ }
}
& 
\makecell{
\scalebox{0.6}{\cite{scozzari_2013} }
}
\\\hline
LB
&  
\makecell{
\scalebox{0.6}{ }
}
& 
\makecell{
\scalebox{0.6}{\cite{strub2019} }
}
\\\hline
LGR
&  
\makecell{
\scalebox{0.6}{\cite{wu2019}; }
}
& 
\makecell{
\scalebox{0.6}{ }
}
\\\hline
KS
&  
\makecell{
\scalebox{0.6}{ }
}
& 
\makecell{
\scalebox{0.6}{\cite{guastaroba_2012}; }
}
\\\hline
VNS
&  
\makecell{
\scalebox{0.6}{ }
}
& 
\makecell{
\scalebox{0.6}{\cite{wu2017_clust} }
}
\\\hline
DA
&  
\makecell{
\scalebox{0.6}{ }
}
& 
\makecell{
\scalebox{0.6}{\cite{stoyan2010} }
}
\\\hline
IWO
&  
\makecell{
\scalebox{0.6}{\cite{affolter2016} }
}
& 
\makecell{
\scalebox{0.6}{ }
}
\\\hline
MOGA
&  
\makecell{
\scalebox{0.6}{\cite{chiam2013}; }
}
& 
\makecell{
\scalebox{0.6}{ }
}
\\\hline
MOPS
&  
\makecell{
\scalebox{0.6}{\cite{zhang2018_proceedings}; }
}
& 
\makecell{
\scalebox{0.6}{ }
}
\\\hline
SA
&  
\makecell{
\scalebox{0.6}{\cite{fernandez-lorenzo2021}; }
}
& 
\makecell{
\scalebox{0.6}{ }
}
\\\hline
HP
&  
\makecell{
\scalebox{0.6}{\cite{fernandez-lorenzo2021}; }
}
& 
\makecell{
\scalebox{0.6}{ }
}
\\\hline
\end{tabular}}
\caption{Heuristic categorization result}
\label{tab:cat_heur}
\end{table}

\begin{figure*}[h!]
\centering
    \scalebox{0.75}{
    \includegraphics[width=\linewidth]{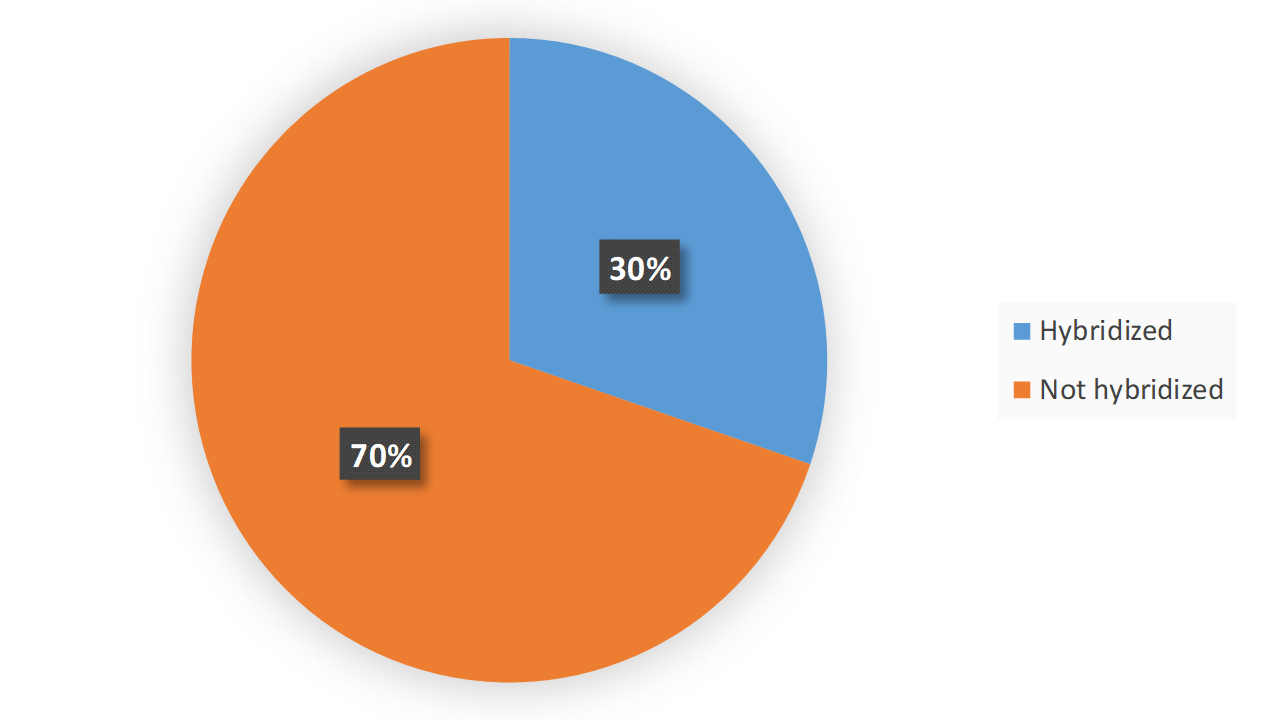}
    }
 \caption{Percentage of works that performed/did not performed hybridization}
 \label{fig:hybrid_vs_nonhybrid}
\end{figure*}

Since there are 13 hybridized heuristics, then the answer to RQ \#8 is positive. Solvers are integrated with heuristics mainly to perform the capital allocation, which eliminates the need for constraint handling approaches and/or repairing mechanisms. Then, in those cases, heuristics were used to perform asset selection only. Table \ref{tab:cat_heur_summ} summarizes the associated quantitative framework subcategory, the number of times that a specific heuristic was applied to a specific model. An interesting result is that all heuristics/metaheuristics applied to MILP (Mixed-Integer Linear Programming) are hybridized. Figure \ref{fig:hybrid_vs_nonhybrid} presents the relative number of works that performed/did not performed hybridization of heuristics/metaheuristics. 

\begin{figure*}[h!]
\centering
\scalebox{0.75}{
    \includegraphics[width=\linewidth]{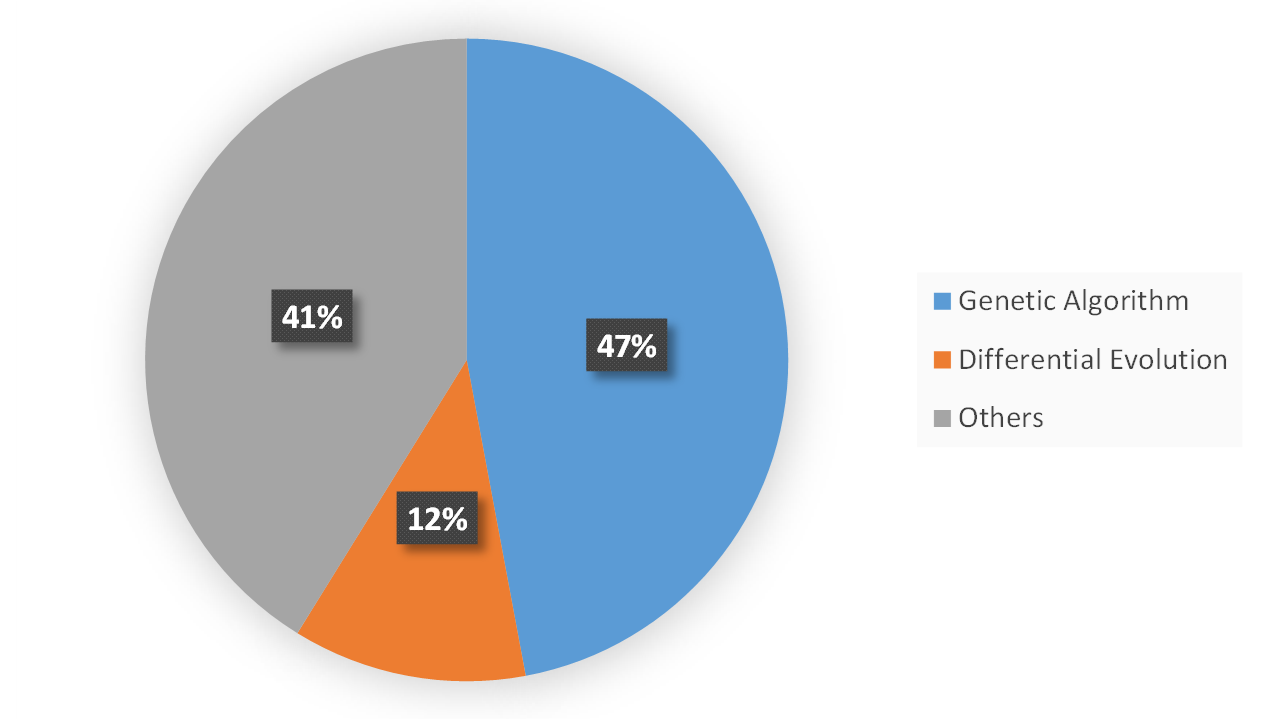}}
 \caption{Percentage of works per heuristic. The two main heuristics were highlighted.}
 \label{fig:mainheur}
\end{figure*}

The answer to RQ \#2 is positive. From Table \ref{tab:cat_heur_summ} it can be observed that the vast amount of the developed heuristics/metaheuristics solutions were applied to mathematical programming formulations more often, more specifically to MINLP (Mixed-Integer Nonlinear Programming) formulations. Although MINLP is the most typical way of applying a heuristic to the problem, MILP formulations may also demand heuristics due to the combinatorial nature of the cardinality constraint. Figure \ref{fig:mainheur} shows the relative number of works per heuristics/metaheuristics. The two main heuristics were highlighted and the other heuristics (applied only once) were clustered as a single group.

\begin{table}[h!]
\centering
\begin{tabular}{@{}cccc@{}}
\toprule
Framework & Algorithm & $N$ & $N_{hybrid}$ 
\\ \midrule
\makecell{Mixed-integer\\Non-linear programming}
&   GA   &   9 & 5 \\
& DE & 4 & 1\\
& TS & 1 & 0\\
& GC & 2 & 0\\
& BEBO & 1 & 0\\
& CS & 1 & 1\\
& LB & 1 & 1\\
& SA & 1 & 0\\
& HP & 1 & 0\\
&   IWO   &  1 & 0 \\
\midrule
\makecell{Mixed-integer\\linear programming}
&   GA   &  1 & 1 \\
&   KS   &  1 & 1 \\
&   VNS   &  1 & 1 \\
\midrule
\makecell{Mixed-integer\\Conic programming}
&   LGR   &  1 & 0 \\
\midrule
Stochastic Programming
&   DA   & 1 & 1 \\
\midrule
\makecell{Multi-objective\\Optimization} 
&   GA   & 1 & 0 \\
&   MOGA   &  1 & 0 \\
&   MOPS   &  1 & 0 \\
\midrule
Cointegration 
&   GA   &  1 & 0 \\
\midrule
\makecell{Regression with\\ regularization} 
&   GA   &  2 & 0 \\
\midrule
Regression 
& GA & 0 & 1 \\
\midrule
\textbf{Total} & & 32 & 13\\
\bottomrule
\end{tabular}
\caption{Heuristic applications summary}
\label{tab:cat_heur_summ}
\end{table}

The answer to RQ \#7 is positive. The works that use heuristic/metaheuristic solution approaches are focused on the application of two heuristics: Genetic and Differential Evolution algorithms.

\subsection{Analysis of heuristic/metaheuristic performance evaluations}

A heuristics/metaheuristics success is indicated by adequate evaluation metrics. In this subsection, all evaluation metrics used to analyze heuristics in the index tracking problem in the last decade were identified. 36 different metrics were found. Then, it was decided to highlight the most used (applied at least 3 times) metrics among the 36. 9 out of 36 metrics were applied at least 3 times, which were: CPU time, GAP (difference between the approximate solution and the best-known solution), Correlation With Respect to the Index (CWRTI), Mean Squared Error (MSE), Information Ratio (IR), Root Mean Squared Error (RMSE), Mean Excess Return (MER), Annualized Return (ANNR) and Annualized Tracking Error (ANNTE). Table \ref{tab:cat_evalmetrics} shows the highlighted metrics and the associated articles.

\begin{table}[h!]
\centering
\scalebox{1.2}{
\begin{tabular}{@{}cc@{}}
\toprule
Metric & Articles 
\\ \midrule
CPU Time
& \multicolumn{1}{l}{
\begin{tabular}[c]{@{}l@{}}
\scalebox{0.8}{\cite{garcia2018}; }
\scalebox{0.8}{\cite{grishina2017}; }\\
\scalebox{0.8}{\cite{mutunge_2018}; }
\scalebox{0.8}{\cite{santanna2017_heur}; }\\
\scalebox{0.8}{\cite{strub2019}; }
\scalebox{0.8}{\cite{wang_2018}; }\\
\scalebox{0.8}{\cite{guastaroba_2012}; }
\scalebox{0.8}{\cite{stoyan2010}}
\end{tabular}
}\\ 
\midrule
GAP
& \multicolumn{1}{l}{
\begin{tabular}[c]{@{}l@{}}
\scalebox{0.8}{\cite{guastaroba_2012}; }
\scalebox{0.8}{\cite{santanna2017_heur}; }\\
\scalebox{0.8}{\cite{scozzari_2013}; }
\scalebox{0.8}{\cite{wu2017_clust}; }
\scalebox{0.8}{\cite{strub2016_proceedings}; }\\
\scalebox{0.8}{\cite{fernandez-lorenzo2021}}
\end{tabular}
}\\ 
\midrule
CWRTI
& \multicolumn{1}{l}{
\begin{tabular}[c]{@{}l@{}}
\scalebox{0.8}{\cite{acosta-gonzalez2015}; } \scalebox{0.8}{\cite{scozzari_2013}; } \\
\scalebox{0.8}{\cite{andriosopoulos2013}; }
\scalebox{0.8}{\cite{andriosopoulos2014}}\\
\end{tabular}
}\\ 
\midrule
MSE
& \multicolumn{1}{l}{
\begin{tabular}[c]{@{}l@{}}
\scalebox{0.8}{\cite{chiam2013}; }
\scalebox{0.8}{\cite{ni2013}; }\\
\scalebox{0.8}{\cite{garcia2018}; }
\scalebox{0.8}{\cite{strub2019}}

\end{tabular}
}\\ 
\midrule
IR
& \multicolumn{1}{l}{
\begin{tabular}[c]{@{}l@{}}
\scalebox{0.8}{\cite{acosta-gonzalez2015}; }
\scalebox{0.8}{\cite{zhang2018_proceedings}; } \\
\scalebox{0.8}{\cite{andriosopoulos2014}; }
\scalebox{0.8}{\cite{giuzio2017}} \\
\end{tabular}
}\\ 
\midrule
RMSE
& \multicolumn{1}{l}{
\begin{tabular}[c]{@{}l@{}}
\scalebox{0.8}{\cite{andriosopoulos2013}; }
\scalebox{0.8}{\cite{andriosopoulos2014}; }\\
\scalebox{0.8}{\cite{santanna2017_heur}; }
\scalebox{0.8}{\cite{chen2011_proceedings}}
\end{tabular}
}\\ 
\midrule
MER
& \multicolumn{1}{l}{
\begin{tabular}[c]{@{}l@{}}
\scalebox{0.8}{\cite{acosta-gonzalez2015}; }\\
\scalebox{0.8}{\cite{andriosopoulos2013}; }
\scalebox{0.8}{\cite{andriosopoulos2014}}\\
\end{tabular}
}\\ 
\midrule
ANNR
& \multicolumn{1}{l}{
\begin{tabular}[c]{@{}l@{}}
\scalebox{0.8}{\cite{acosta-gonzalez2015}; }\\
\scalebox{0.8}{\cite{andriosopoulos2013}; }
\scalebox{0.8}{\cite{andriosopoulos2014}}\\
\end{tabular}
}\\
\midrule
ANNTE
& \multicolumn{1}{l}{
\begin{tabular}[c]{@{}l@{}}
\scalebox{0.8}{\cite{acosta-gonzalez2015}; }\\
\scalebox{0.8}{\cite{giuzio2017}; }
\scalebox{0.8}{\cite{scozzari_2013}}
\end{tabular}
}\\
\bottomrule
\end{tabular}}
\caption{Evaluation metrics summary}
\label{tab:cat_evalmetrics}
\end{table}

\begin{figure*}[h!]
\centering
\scalebox{0.75}{
    \includegraphics[width=\linewidth]{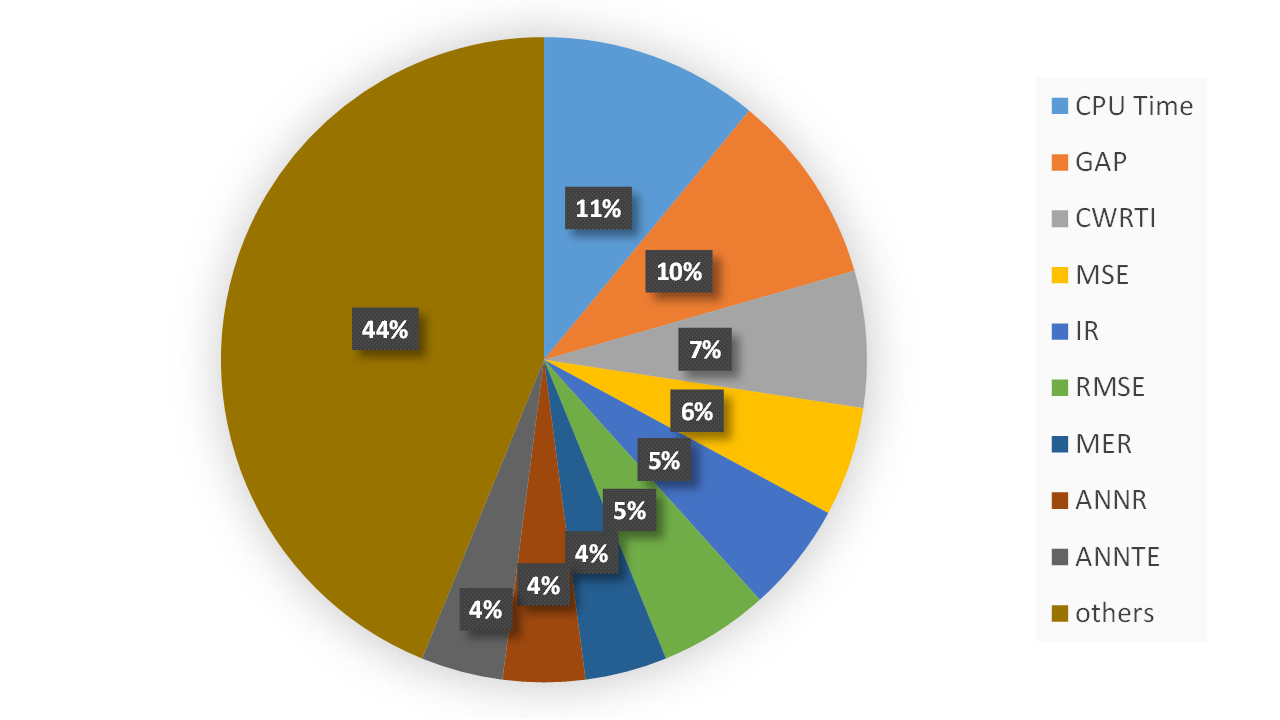}}
 \caption{Percentage of works per evaluation metrics. The nine main evaluation metrics were highlighted.}
 \label{fig:mainevalmetrics}
\end{figure*}

Figure \ref{fig:mainevalmetrics} presents the percentage of works that applied each of the nine highlighted metrics and the other 36 metrics. The answer to RQ \#9 is positive since those nine metrics are prevalent among the other 36 metrics.

To answer RQ \#10, a search about comparison types was performed. The identified comparison types were: Heuristic against Heuristic, Heuristic against CPLEX, Heuristic against Gurobi, Heuristic against Projected Gradient algorithm, Heuristic against the interior-point algorithm, Heuristic against Cyclic Coordinate Descent algorithm and Heuristic against Random Selection. Table \ref{tab:cat_comparisons} shows the comparison types and related papers.

\begin{table}[h!]
\centering
\scalebox{1.2}{
\begin{tabular}{@{}cc@{}}
\toprule
Comparison Type & Articles 
\\ \midrule
\makecell{Heuristic Vs Heuristic}
& \multicolumn{1}{l}{
\begin{tabular}[c]{@{}l@{}}
\scalebox{0.8}{\cite{acosta-gonzalez2015}; }
\scalebox{0.8}{\cite{andriosopoulos2013}; } \\
\scalebox{0.8}{\cite{andriosopoulos2014}; }
\scalebox{0.8}{\cite{chiam2013}; } \\
\scalebox{0.8}{\cite{garcia2018}; }
\scalebox{0.8}{\cite{grishina2017}; } \\
\scalebox{0.8}{\cite{ni2013}; }
\scalebox{0.8}{\cite{wu2017_clust}; } \\
\scalebox{0.8}{\cite{strub2016_proceedings}; \cite{fernandez-lorenzo2021}} 
\end{tabular}
}\\ 
\midrule
\makecell{Heuristic Vs CPLEX}
& \multicolumn{1}{l}{
\begin{tabular}[c]{@{}l@{}}
\scalebox{0.8}{\cite{mutunge_2018}; } \\
\scalebox{0.8}{\cite{santanna2017_heur}; } \\
\scalebox{0.8}{\cite{scozzari_2013}; }
\scalebox{0.8}{\cite{stoyan2010}; }
\end{tabular}
}\\ 
\midrule
\makecell{Heuristic Vs Gurobi}
& \multicolumn{1}{l}{
\begin{tabular}[c]{@{}l@{}}
\scalebox{0.8}{\cite{strub2019}; }
\scalebox{0.8}{\cite{wang_2018}; } \\
\scalebox{0.8}{\cite{strub2016_proceedings}; } 
\end{tabular}
}\\ 
\midrule
\makecell{Heuristic Vs ProjGrad}
& \multicolumn{1}{l}{
\begin{tabular}[c]{@{}l@{}}
\scalebox{0.8}{\cite{giuzio2017}; }
\scalebox{0.8}{\cite{xu2016}; }
\end{tabular}
}\\ 
\midrule
\makecell{Heuristic Vs IntPoint}
& \multicolumn{1}{l}{
\begin{tabular}[c]{@{}l@{}}
\scalebox{0.8}{\cite{giuzio2017}; }
\end{tabular}
}\\ 
\midrule
\makecell{Heuristic Vs CycD}
& \multicolumn{1}{l}{
\begin{tabular}[c]{@{}l@{}}
\scalebox{0.8}{\cite{giuzio2017}; }
\end{tabular}
}\\ 
\midrule
\makecell{Heuristic Vs Random}
& \multicolumn{1}{l}{
\begin{tabular}[c]{@{}l@{}}
\scalebox{0.8}{\cite{affolter2016}; }
\scalebox{0.8}{\cite{strub2016_proceedings}; } 
\end{tabular}
}\\ 
\bottomrule
\end{tabular}}
\caption{Comparison types summary}
\label{tab:cat_comparisons}
\end{table}

\begin{figure*}[h!]
\centering
\scalebox{0.75}{
    \includegraphics[width=\linewidth]{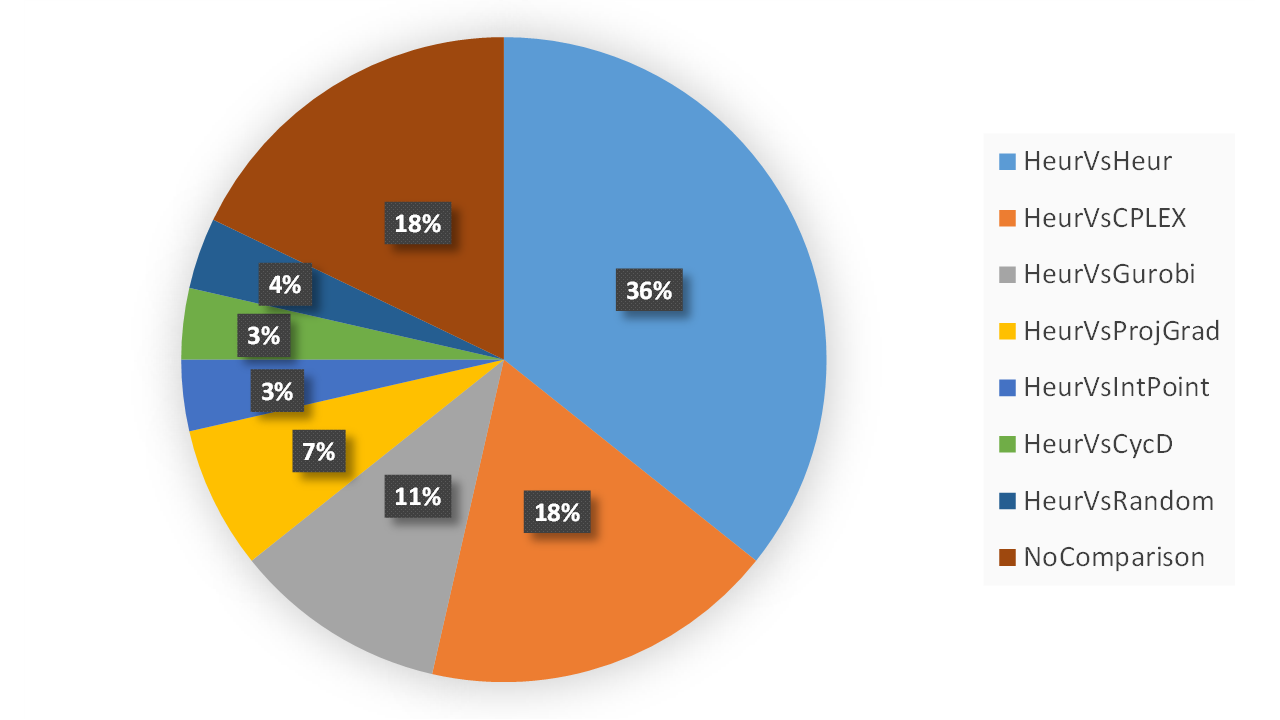}}
 \caption{Percentage of works per comparison type.}
 \label{fig:comparisons}
\end{figure*}

Figure \ref{fig:comparisons} presents the percentage of articles that performed each type of comparison. It can be observed that the answer to RQ \#10 is positive since most of the authors that developed a heuristic/metaheuristic compared it against another heuristic/metaheuristic or the commercial solver CPLEX.

\subsection{Analysis of index tracking models solved by metaheuristics/heuristics}

The objective functions found were: Root Mean Squared Error (RMSE), Mean Squared Error (MSE), Mean Absolute Deviation (MAD), Absolute Deviation (AD), Tracking Error Variance (TEV), Sum of Errors Squares (SES), Correlation, Mean Index Excess Return (MIER), Accumulated Excess Return (AER), Augmented Dickey-Fuller \textit{t} statistic (ADF), return,  utility, penalties (on tracking error, portfolio size, capital, and volatility), Value at Risk (VaR), Sensibility, and Transaction Costs (TC). The classification of each article in an objective function is presented in table \ref{tab:cat_obj_summ}. From this table we can see that the mathematical programming framework contains a more diverse set of objective functions when compared to others. Also, the possibility of using multi-objective formulations makes the mathematical framework more interesting when the analyst wants to consider the inclusion of investor preferences w.r.t the considered objectives.

\begin{table}[h!]
\centering
\scalebox{0.7}{
\begin{tabular}{@{}ccc@{}}
\toprule
Framework & Objective & Articles 
\\ \midrule
\makecell{Mixed-integer\\Non-linear programming}
&   RMSE   
& \multicolumn{1}{l}{
\begin{tabular}[c]{@{}l@{}}
\scalebox{0.8}{\cite{andriosopoulos2014}; }\\
\scalebox{0.8}{\cite{andriosopoulos2013}; }\\
\end{tabular}
}\\ \\
&   MSE   
& \multicolumn{1}{l}{
\begin{tabular}[c]{@{}l@{}}
\scalebox{0.8}{\cite{garcia2018}; }\\
\scalebox{0.8}{\cite{santanna2017_heur}; }\\
\scalebox{0.8}{\cite{scozzari_2013}; }\\
\scalebox{0.8}{\cite{strub2019}; }\\
\scalebox{0.8}{\cite{fernandez-lorenzo2021}; }\\
\end{tabular}
}\\ \\
&   MAD  
& \multicolumn{1}{l}{
\begin{tabular}[c]{@{}l@{}}
\scalebox{0.8}{\cite{grishina2017}; }\\
\end{tabular}
}\\ \\
&   SES 
&  \multicolumn{1}{l}{
\begin{tabular}[c]{@{}l@{}}
\scalebox{0.8}{\cite{wang_2018}; }\\
\scalebox{0.8}{\cite{xu2016}; }\\
\end{tabular}
}\\ \\
&   TEV   
& \multicolumn{1}{l}{
\begin{tabular}[c]{@{}l@{}}
\scalebox{0.8}{\cite{mutunge_2018}; }\\
\scalebox{0.8}{\cite{strub2016_proceedings}; }\\
\end{tabular}
} \\ \\
&   Return   
& \multicolumn{1}{l}{
\begin{tabular}[c]{@{}l@{}}
\scalebox{0.8}{\cite{affolter2016}}\\
\end{tabular}
}\\ \\
&   Penalty$_{volatility}$   
& \multicolumn{1}{l}{
\begin{tabular}[c]{@{}l@{}}
\scalebox{0.8}{\cite{affolter2016}}\\
\end{tabular}
}\\ \\
&   Penalty$_{portfolioSize}$   
& \multicolumn{1}{l}{
\begin{tabular}[c]{@{}l@{}}
\scalebox{0.8}{\cite{affolter2016}}\\
\end{tabular}
}\\ \\
&   Penalty$_{capital}$   
& \multicolumn{1}{l}{
\begin{tabular}[c]{@{}l@{}}
\scalebox{0.8}{\cite{affolter2016}}\\
\end{tabular}
}\\ \\
&   Penalty$_{TE}$   
& \multicolumn{1}{l}{
\begin{tabular}[c]{@{}l@{}}
\scalebox{0.8}{\cite{affolter2016}}\\
\end{tabular}
}\\ 
\midrule
\makecell{Mixed-integer\\linear programming}
&   MAD 
&  \multicolumn{1}{l}{
\begin{tabular}[c]{@{}l@{}}
 \scalebox{0.8}{\cite{wang2012_main}}
\end{tabular}
}\\ \\
&  AD   
&  \multicolumn{1}{l}{
\begin{tabular}[c]{@{}l@{}}
\scalebox{0.8}{\cite{guastaroba_2012};}
\end{tabular}
}\\ \\
&  Correlation  
&  \multicolumn{1}{l}{
\begin{tabular}[c]{@{}l@{}}
\scalebox{0.8}{\cite{wu2017_clust};}
\end{tabular}
}\\ 
\midrule
\makecell{Mixed-integer\\Conic programming}
&   MIER    
&  \multicolumn{1}{l}{
\begin{tabular}[c]{@{}l@{}}
\scalebox{0.8}{\cite{wu2019}; }\\
\end{tabular}
}\\ 
\midrule
Stochastic Programming
&  AD   
&  \multicolumn{1}{l}{
\begin{tabular}[c]{@{}l@{}}
\scalebox{0.8}{\cite{stoyan2010};}
\end{tabular}
}\\ 
\midrule
\makecell{Multi-objective\\Optimization} 
&   RMSE  
&  \multicolumn{1}{l}{
\begin{tabular}[c]{@{}l@{}}
\scalebox{0.8}{\cite{ni2013}; }\\
\scalebox{0.8}{\cite{zhang2018_proceedings}; }\\
\end{tabular}
}\\ \\
&   MSE  
&  \multicolumn{1}{l}{
\begin{tabular}[c]{@{}l@{}}
\scalebox{0.8}{\cite{chiam2013}; }\\
\end{tabular}
}\\ \\
&   AER  
&  \multicolumn{1}{l}{
\begin{tabular}[c]{@{}l@{}}
\scalebox{0.8}{\cite{ni2013}; }\\
\end{tabular}
}\\ \\
&   TC 
&  \multicolumn{1}{l}{
\begin{tabular}[c]{@{}l@{}}
\scalebox{0.8}{\cite{chiam2013}; }\\
\end{tabular}
}\\ \\
&   SBTY 
&  \multicolumn{1}{l}{
\begin{tabular}[c]{@{}l@{}}
\scalebox{0.8}{\cite{zhang2018_proceedings}; }\\
\end{tabular}
}\\ \\
&   VaR
&  \multicolumn{1}{l}{
\begin{tabular}[c]{@{}l@{}}
\scalebox{0.8}{\cite{zhang2018_proceedings}; }\\
\end{tabular}
}\\ 
\midrule
Cointegration 
&   ADF
&  \multicolumn{1}{l}{
\begin{tabular}[c]{@{}l@{}}
\scalebox{0.7}{\cite{acosta-gonzalez2015}; }\\
\end{tabular}
}\\ \\
&   Correlation
&  \multicolumn{1}{l}{
\begin{tabular}[c]{@{}l@{}}
\scalebox{0.7}{\cite{acosta-gonzalez2015}; }\\
\end{tabular}
}\\ 
\midrule
\makecell{Regression} 
&   RMSE
&  \multicolumn{1}{l}{
\begin{tabular}[c]{@{}l@{}}
\scalebox{0.8}{\cite{chen2011_proceedings}; }\\
\end{tabular}
}\\ 
\midrule
\makecell{Regression with \\regularization} 
&   SES
&  \multicolumn{1}{l}{
\begin{tabular}[c]{@{}l@{}}
\scalebox{0.8}{\cite{xu2016}; }\\
\end{tabular}
}\\ \\
&   MSE
&  \multicolumn{1}{l}{
\begin{tabular}[c]{@{}l@{}}
\scalebox{0.8}{\cite{giuzio2017}; }\\
\end{tabular}
}\\ 
\bottomrule
\end{tabular}
}
\caption{Objective function occurrence summary}
\label{tab:cat_obj_summ}
\end{table}

\begin{figure*}[h!]
\centering
\scalebox{0.75}{
    \includegraphics[width=\linewidth]{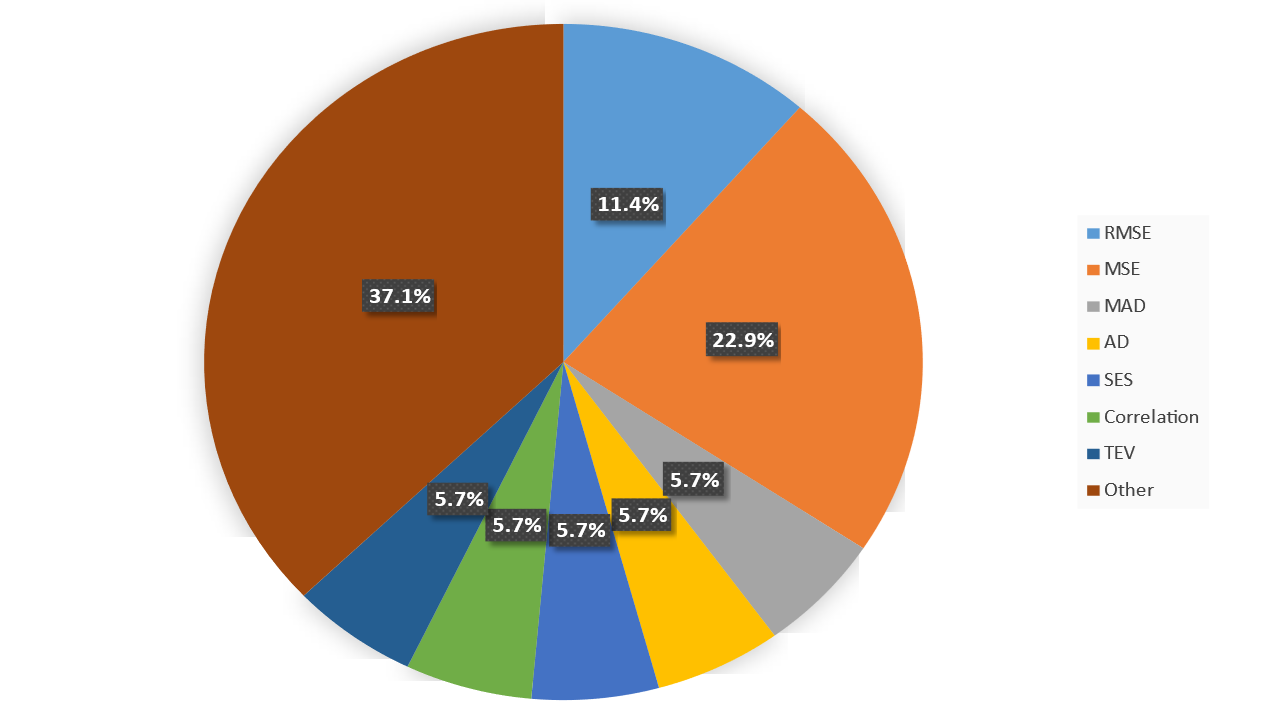}}
 \caption{Percentage of works per objective function.}
 \label{fig:obj_func}
\end{figure*}

The relative frequency of works per objective function is shown in Figure \ref{fig:obj_func}. The answer to RQ \#11 is positive since a good part of the works adopts RMSE and MSE.

\begin{figure*}[h!]
\centering
\scalebox{0.75}{
    \includegraphics[width=\linewidth]{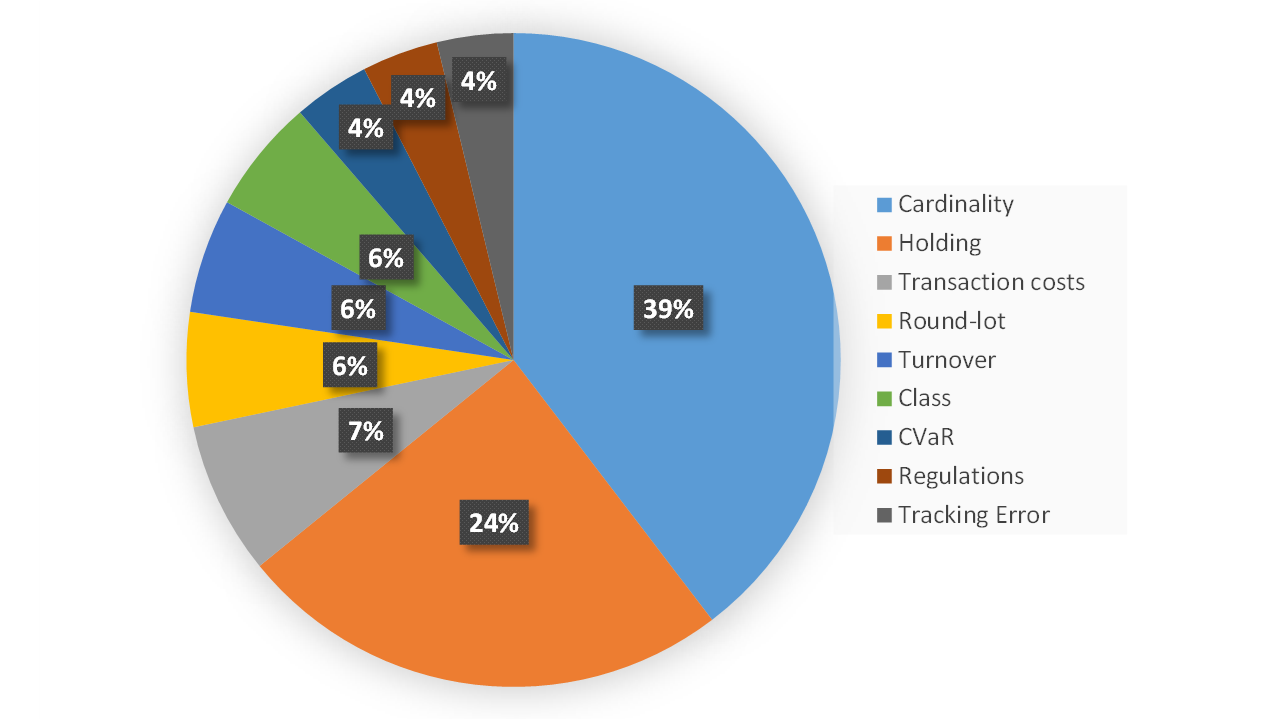}}
 \caption{Percentage of works per constraint.}
 \label{fig:constraints}
\end{figure*}

Two main constraints are usually applied to index tracking problems: cardinality and holding. These constraints were adopted by articles that used mathematical programming only since their use is obligatory for the index tracking model in this framework. In this work, more practical constraints were taken into account to answer RQ \#12. The constraints found were: Transaction Costs, Tracking Error, Turnover, Market Regulations, Class, CVaR and Round-lot. As well as the two main constraints, these practical constraints only occurred in the mathematical programming modelling framework. This is because the mathematical programming framework is more flexible in terms of inclusion of practical constraints, which is not the case for other modelling frameworks.
The relative number of works per constraint is shown in Figure \ref{fig:constraints}. 

Since the constraints are well distributed among the articles, there is no prevalence of a specific practical constraint when applying heuristics/metaheuristics and the answer relative to RQ \#12 is negative. Table \ref{tab:cat_constr_summ} shows the mathematical programming subcategory, constraint name and associated articles.

\begin{table}[h!]
\centering
\scalebox{0.96}{
\begin{tabular}{@{}ccc@{}}
\toprule
Framework & Constraint & Articles 
\\ \midrule
\makecell{Mixed-integer\\Non-linear programming}
&   Tracking Error    
& \multicolumn{1}{l}{
\begin{tabular}[c]{@{}l@{}}
\scalebox{0.8}{\cite{santanna2017_heur}; }\\
\end{tabular}
}\\ \\
&   Turnover    
&  \multicolumn{1}{l}{
\begin{tabular}[c]{@{}l@{}}
\scalebox{0.8}{\cite{scozzari_2013}; }\\
\end{tabular}
}\\ \\
&   Market regulation    
&  \multicolumn{1}{l}{
\begin{tabular}[c]{@{}l@{}}
\scalebox{0.8}{\cite{scozzari_2013};}\\
\scalebox{0.8}{\cite{strub2019};}\\
\end{tabular}
}\\ \\
&  Class    
&  \multicolumn{1}{l}{
\begin{tabular}[c]{@{}l@{}}
\scalebox{0.8}{\cite{wang_2018};}
\end{tabular}
}\\ \\
&   CVaR    
& \multicolumn{1}{l}{
\begin{tabular}[c]{@{}l@{}}
\scalebox{0.8}{\cite{wang_2018}; }\\
\end{tabular}
}\\ \\
&   Round-lot
& \multicolumn{1}{l}{
\begin{tabular}[c]{@{}l@{}}
\scalebox{0.8}{\cite{wang_2018}; }\\
\end{tabular}
}\\ 
\midrule
\makecell{Mixed-integer\\linear programming}
&   Transaction costs   
&  \multicolumn{1}{l}{
\begin{tabular}[c]{@{}l@{}}
\scalebox{0.8}{\cite{guastaroba_2012};}\\ \scalebox{0.8}{\cite{wu2017_clust}}
\end{tabular}
}\\ \\
&  Class    
&  \multicolumn{1}{l}{
\begin{tabular}[c]{@{}l@{}}
\scalebox{0.8}{\cite{wu2017_clust};}
\end{tabular}
}\\ \\
&   CVaR    
& \multicolumn{1}{l}{
\begin{tabular}[c]{@{}l@{}}
\scalebox{0.8}{\cite{wang2012_main}; }\\
\end{tabular}
}\\ 
\midrule
\makecell{Mixed-integer\\Conic programming}
&   Tracking Error    
&  \multicolumn{1}{l}{
\begin{tabular}[c]{@{}l@{}}
\scalebox{0.8}{\cite{wu2019}; }\\
\end{tabular}
}\\ 
\midrule
Stochastic Programming
&   Turnover    
&  \multicolumn{1}{l}{
\begin{tabular}[c]{@{}l@{}}
\scalebox{0.8}{\cite{stoyan2010}; }\\
\end{tabular}
}\\ \\
&  Class    
&  \multicolumn{1}{l}{
\begin{tabular}[c]{@{}l@{}}
\scalebox{0.8}{\cite{stoyan2010};}
\end{tabular}
}\\ 
\midrule
\makecell{Multi-objective\\Optimization} 
&   Transaction costs   
&  \multicolumn{1}{l}{
\begin{tabular}[c]{@{}l@{}}
\scalebox{0.8}{\cite{ni2013}; }\\
\scalebox{0.8}{\cite{zhang2018_proceedings}; }\\
\end{tabular}
}\\ \\
&   Turnover   
&  \multicolumn{1}{l}{
\begin{tabular}[c]{@{}l@{}}
\scalebox{0.8}{\cite{ni2013}; }\\
\end{tabular}
}\\ \\
&   Round-lot  
&  \multicolumn{1}{l}{
\begin{tabular}[c]{@{}l@{}}
\scalebox{0.8}{\cite{chiam2013}; }\\
\end{tabular}
}\\
\bottomrule
\end{tabular}}
\caption{Practical constraint occurrence summary}
\label{tab:cat_constr_summ}
\end{table}

Even though this literature review focuses on the characteristics of models solved by heuristics/metaheuristics, the results presented in this part of the review may also be useful for researchers that are developing index tracking models. They can compare their objective functions and practical constraints with other authors' models that also included them in a specific quantitative modelling framework.

\subsection{Categorization - data sources for index tracking problems solved by heuristic/metaheuristic}

To answer research question RQ \#13 it was necessary to identify all data sources. The following data sources were identified: OR-library, Datastream, Historical Stock Data Downloader (HSDD), Economatica, Bloomberg, Markit, Academic research lab (University) and Yahoo Finance. Table \ref{tab:cat_databases} shows all the databases and associated articles.

\begin{table}[h!]
\centering
\scalebox{1.2}{
\begin{tabular}{@{}cc@{}}
\toprule
Data source & Articles 
\\ \midrule
Datastream
& \multicolumn{1}{l}{
\begin{tabular}[c]{@{}l@{}}
\scalebox{0.8}{\cite{affolter2016}; }
\scalebox{0.8}{\cite{andriosopoulos2013}; }\\
\scalebox{0.8}{\cite{andriosopoulos2013}; }
\scalebox{0.8}{\cite{strub2019}; }\\
\end{tabular}
}\\ 
\midrule
OR-library
& \multicolumn{1}{l}{
\begin{tabular}[c]{@{}l@{}}
\scalebox{0.8}{\cite{garcia2018}; }
\scalebox{0.8}{\cite{grishina2017}; }\\
\scalebox{0.8}{\cite{mutunge_2018}; }
\scalebox{0.8}{\cite{strub2019}; }\\
\scalebox{0.8}{\cite{guastaroba_2012}; }
\scalebox{0.8}{\cite{wang2012_main}; }\\
\scalebox{0.8}{\cite{chiam2013}; }
\scalebox{0.8}{\cite{strub2016_proceedings}; }
\end{tabular}
}\\ 
\midrule
HSDD
& \multicolumn{1}{l}{
\begin{tabular}[c]{@{}l@{}}
\scalebox{0.8}{\cite{mutunge_2018}; }
\end{tabular}
}\\ 
\midrule
Economatica
& \multicolumn{1}{l}{
\begin{tabular}[c]{@{}l@{}}
\scalebox{0.8}{\cite{santanna2017_heur}; }
\end{tabular}
}\\ 
\midrule
Bloomberg
& \multicolumn{1}{l}{
\begin{tabular}[c]{@{}l@{}}
\scalebox{0.8}{\cite{santanna2017_heur}; }
\end{tabular}
}\\ 
\midrule
Markit
& \multicolumn{1}{l}{
\begin{tabular}[c]{@{}l@{}}
\scalebox{0.8}{\cite{wu2019}; }
\end{tabular}
}\\ 
\midrule
University
& \multicolumn{1}{l}{
\begin{tabular}[c]{@{}l@{}}
\scalebox{0.8}{\cite{wu2017_clust}; }
\end{tabular}
}\\ 
\midrule
Yahoo Finance
& \multicolumn{1}{l}{
\begin{tabular}[c]{@{}l@{}}
\scalebox{0.8}{\cite{acosta-gonzalez2015}; }
\end{tabular}
}\\
\bottomrule
\end{tabular}
}
\caption{Data sources summary}
\label{tab:cat_databases}
\end{table}

\begin{figure*}[h!]
\centering
\scalebox{0.75}{
    \includegraphics[width=\linewidth]{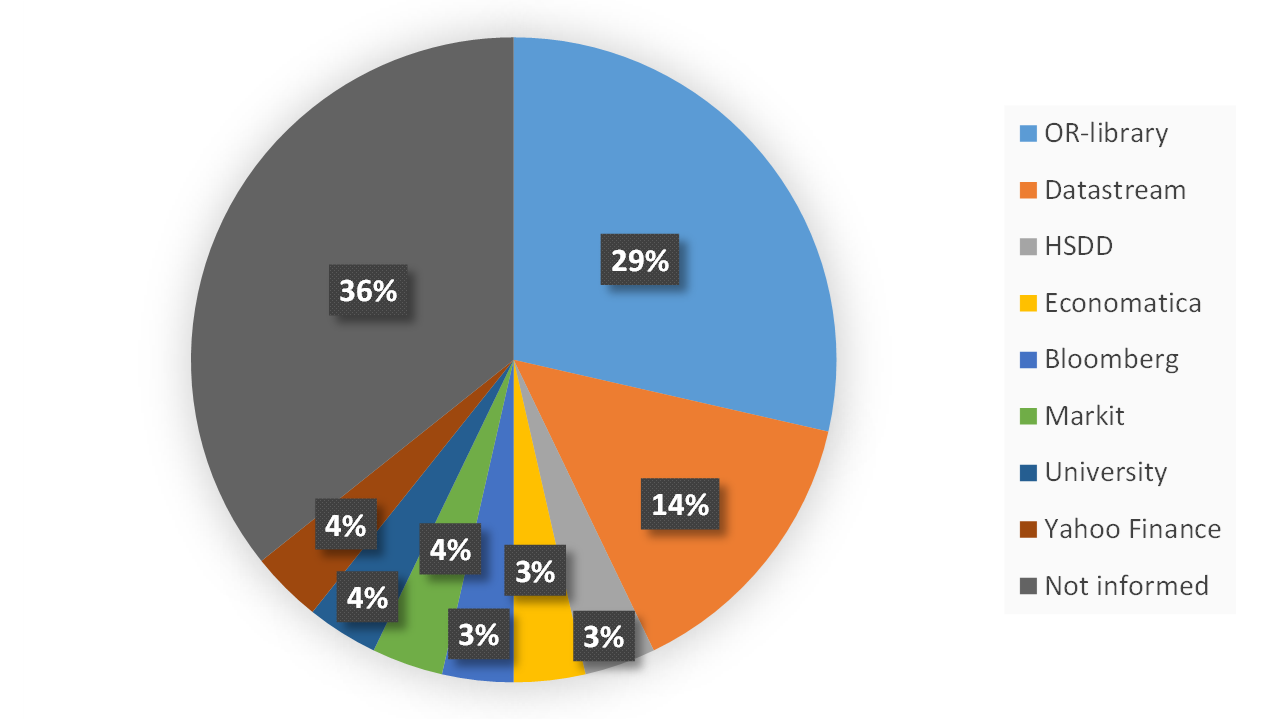}}
 \caption{Percentage of works per database.}
 \label{fig:databases}
\end{figure*}

Figure \ref{fig:databases} shows the percentage of works per data source. OR-library and Datastream were adopted by most of the works that are transparent about their databases. Then, the answer to RQ \#13 is OR-library and Datastream.

\section{Discussion}
The objective of this systematic review was to develop an investigation concerning the current solution approaches for the index tracking problem using a set of research questions. The methodology consisted of searching for articles and conference papers written in the English language related to this class of portfolio selection problems. The set of research questions was divided into two parts. 

The first part refers to general solution approaches for the index tracking problem and comparison among two groups: heuristic and non-heuristic methods. The production of both approaches grew in the last decade, also, non-heuristic methods were more adopted than their counterparts. On the other hand, heuristics obtained the best performance when citation impact is taken into consideration, considering the metrics adopted in this work.

The second part refers to a specific analysis of heuristic/metaheuristic approaches applications developed for this problem. The first part of this analysis consisted of investigating the main heuristics, if there were hybridized heuristics/metaheuristics, comparison against other methods, and the associated evaluation metrics. Next, the model structure of the models solved by approximated methods was studied, taking into consideration objective functions, constraints, and data used in the problem. Table \ref{tab:rqs_idx_smmry} summarizes the answers to each research question.

Researchers can consider this work for further developments in the index tracking problems, especially if they're using approximate solution methods for solving these. Also, information about the state-of-the-art model structure is detailed for studies considering model development for this class of problems.

\begin{table*}[h!]
\centering
\scalebox{1.0}{
\begin{tabular}{cc}
\hline
RQ  & Findings                                                                                                                                \\ \hline

\#1 & \makecell{Index tracking solution methods are more relevant to journals focusing \\on operations research and computer science }                                                            \\
\#2 & \makecell{The vast amount of the developed heuristics/metaheuristics solutions \\were applied to mathematical programming formulations more often} 
\\
\#3 & \makecell{There has been a growth in the number of non-heuristic methods applied to \\ the index tracking problem}                                                            \\
\#4 & \makecell{There has been a growth in the number of heuristics/metaheuristics applied \\ to the index tracking problem}                                                           \\
\#5 & \makecell{Heuristic approaches are not more used than non-heuristic approaches for \\index tracking problems}                                                            \\
\#6 & \makecell{Heuristic approaches have more cite impact than non-heuristic \\ approaches for index tracking problems}   
\\
\#7 & \makecell{There is a prevalence of using Differential Evolution and Genetic algorithms\\  in index tracking problems}
\\
\#8 & \makecell{Solvers are integrated with heuristics. A total of 11 hybridized\\ heuristics were found}
\\
\#9 & \makecell{There is a prevalence of using specific evaluation metrics for heuristic\\ approaches. The most used metrics were RMSE and MSE}
\\
\#10 & \makecell{Yes heuristics are more compared against other heuristics \\or the CPLEX solver}
\\
\#11 & \makecell{There is a prevalence of solving for a specific tracking error objective \\function when using heuristic approaches. A good part of the \\works adopted RMSE and MSE}
    \\
\#12 & \makecell{No. There is no prevalence of solving for specific practical constraints when\\ using heuristic approaches}
    \\
\#13 & \begin{tabular}[c]{@{}c@{}}The most used databases were OR-library and datastream
    \end{tabular} \\ \hline
\end{tabular}
}
\caption{Summary of the answers to each research question of the index tracking systematic literature review}
\label{tab:rqs_idx_smmry}
\end{table*}

\section{Conclusion}
With the growth of heuristic and non-heuristic solution approaches applied to the index tracking problem in the last decade, journals encompassing operational research and computer science areas were the most promising destinations for research papers concerning this problem. Despite their relatively low production rate in the last decade, heuristics/metaheuristics have more citation impact than non-heuristic solution approaches. Also, it was possible to verify that heuristics/metaheuristics solution approaches are concentrated in the MINLP quantitative modelling framework.

Most of the papers that adopted heuristics/metaheuristics as a solution approach implemented genetic and differential evolution algorithms to their index tracking formulation. Not only pure heuristics were applied to this problem, but also hybridizations using both heuristics and commercial solvers. CPU time and correlation with respect to the index were some of the most used evaluation metrics to indicate heuristics/metaheuristics success. These evaluation metrics were used most often when comparing different heuristics/metaheuristics or when using the CPLEX solver as the main benchmark solution method.

A diversity of objective functions was found among works in index tracking models, such as minimization of tracking error measures, transaction costs minimization and return maximization. A good part of the objective functions used in the formulations solved by heuristics solution approaches consisted of two tracking error measures: Root Mean Squared Error and Mean Squared Error. Although different types of practical constraints were used in index tracking problems solved by heuristic methods in the last decade, such as round-lot and CVaR (Conditional Value at Risk), there is no prevalence of using any of these. The mathematical programming framework contained a more diverse set of objectives and constraints, which makes it more flexible and extensible than other quantitative modelling frameworks. With respect to databases adopted, the most used is the OR-library followed by Datastream. The former database contains public data and the latter is a commercial database.

Future challenges involve the development of index tracking models and heuristics that deal with more practical constraints, such as transactions costs and those involving market regulations. Also, future work should compare the performance of hybrid heuristics and non-hybrid heuristics in index tracking models to evaluate if it is more efficient to adopt a general-purpose solver than a constraint handling or repairing mechanism to deal with the capital allocation constraints. With more machine/deep learning techniques being applied to this problem, future developments may explore how to incorporate more practical constraints in index tracking problems combining these learning techniques with mathematical programming. The mathematical programming framework also enables the multiple objective setting, which has not been much explored in index tracking, especially for many-objectives formulations and interactive multi-objective optimization.

\section*{Acknowledgments}
This work was partially supported by CNPQ (315245/2020-4) and CAPES (001) (Brazilian research agencies), for which we are grateful. 




\bibliographystyle{unsrt}
\bibliography{main}

\end{document}